# Interaction of moiré-induced quantum Hall channels in a locally gated graphene junction

Won Beom Choi[1,2], Myungjin Jeon[1,2], K. Watanabe[3], T. Taniguchi[4], Joonho Jang[1,2]*

[1] Department of Physics and Astronomy, and Institute of Applied Physics, Seoul National University, Seoul 08826, Korea

[2] Center for Correlated Electron Systems, Institute for Basic Science, Seoul 08826, Korea

[3] Research Center for Electronic and Optical Materials, National Institute for Materials Science, 1-1 Namiki, Tsukuba 305-0044, Japan

[4] Research Center for Materials Nanoarchitectonics, National Institute for Materials Science, 1-1 Namiki, Tsukuba 305-0044, Japan

*Corresponding author's e-mail: joonho.jang@snu.ac.kr

**Abstract**

Manipulating electron quantum 1D channels is an important element in the field of quantum information due to their ballistic and phase coherence properties. In GaAs and graphene based two dimensional gas systems, these edge channels have been investigated with both integer and fractional quantum Hall effects, contributing to the realization of electron interferometer and anyon braiding. Often, at the p-n junction in the quantum Hall (QH) regime, the presence of a depletion region due to a band gap or the formation of gaps between the zeroth Landau levels (zLL) suppresses interaction between the co-propagating edge channels of opposing doping regimes and helps to preserve the phase coherence of the channels. Here, we observe a new type of p-n junction in hexagonal boron nitride aligned graphene that lacks both the zLL and band gap. In this system, a van Hove singularity (vHS) emerges at the p-n junctions under magnetic fields of several Tesla, owing to the doping inversion near the secondary Dirac point. By fabricating devices with independently tunable global bottom and local top gates, we enable the study of interactions between p-type and n-type QH edge channels through magnetic breakdown associated with the vHS. These findings provide valuable insights into the interactions of superlattice-induced QH edge channels in hBN-aligned graphene.

# 1. Introduction

Controlling the Quantum Hall (QH) edge - topologically protected and ballistic electron channels - has become crucial due to their potential applications in quantum information. In a two-dimensional gas (2DEG) system based on GaAs, QH interferometers with quantum point contact geometries have enabled the observation of anyonic braiding statistics [1–3], which can be utilized as the topological quantum computing [4]. The physical realization of 2DEG systems is essential for studying QH edge phenomena, and graphene emerged as a promising alternative to realize QH interferometers for its high mobility and stability [5–7]. Fully utilizing the graphene as a building block for quantum computing based on the ballistic channels requires a fundamental understanding of QH edge interactions.

To realize the quantum devices based on high-quality graphene 2DEG, implementing the local control of doping levels and chemical potential variation in nanoscale is necessary. With the action of local gates deposited on hBN, high-quality hBN-encapsulated graphene devices provide robust junctions with *smooth* interfaces, allowing the investigation of the equilibration phenomena at the interfaces among QH edge channels depending on the number of edges propagating along the junction interface [8–13]. For example, in a unipolar configuration, such as N-N-N or P-P-P junctions, when the local-gate area hosts fewer edge channels than the global area, local edge channels can transmit without reflection, a regime called edge transmission (ET). Conversely, when the number of edges in the local-gate area exceeds that in the global area, the reflected edge channels appear in the local-gate area, leading to fractional quantum resistance upon a measurement configuration, a regime called partial equilibration (PE). In the bipolar case, such as the P-N-P or N-P-N junction, QH edge channels in p- and n-doping areas co-propagate, but the interactions between QH edge channels are suppressed due to the presence of a depletion region or a zeroth Landau level (zLL) at the P-N junction, resulting in the effective isolation. Then, sharp corners of the interfaces, such as the intersection of a physical edge and a P-N junction, can be used to induce local equilibration between channels, acting as an electronic beam splitter, allowing interferometric measurements in the QH regime [12,14–16].

Despite the high quality and flexibility offered by van der Waals 2DEG systems, which allow stacking and twisting of multiple layers, previous studies on the QH edge channels and their interactions have been limited to intrinsic monolayer and bilayer graphenes. In particular, a system of graphene with hexagonal boron nitride (hBN) aligned with a very small twisting angle constitutes one of the simplest graphene moiré systems; the isostructural graphene and hBN, with a lattice constant mismatch of approximately 1.8%, induce a moiré potential that modifies the band structure of monolayer graphene, opening band gaps at the first Dirac point (FDP), the hole-doped second Dirac point (hSDP), and the electron-doped second Dirac point (eSDP) [17–22] and producing low-energy van Hove singularities (vHS) [23–25] in the otherwise gapless graphene. The moiré potential further enriches the electronic spectra in magnetic fields, giving rise to complex but systematic changes of Hall quantized conductance, called Hofstadter butterfly, in magnetotransport measurements [17,18,20], suggesting strong modification of QH edge channels under the moiré potential. Investigating these QH edge channels and their interactions could broaden the understanding of the emergent 1D channels in moiré systems.

In this paper, we present high quality hBN-aligned graphene devices with a local top gate and a global bottom gate under high magnetic fields and study the interactions between moiré-induced QH edges. We independently dope the area under the local-gate and the global area of a sample by modifying the chemical potentials, and are able to tune to different types of QH edge channels. Based on our analysis, two qualitatively different types of p-n junctions are clearly distinguished; one with the zLL and an insulating region and the other without them. The observation signifies the potential to realize quantum devices with emergent functionality based on the superlattice-induced electronic band modifications.

## 2. Device fabrication

Our device is fabricated by the dry transfer method, to preserve the high quality interfaces of hBN and graphene heterostructures [26]. We exfoliate thin hBN flakes and a monolayer graphene on a $Si/SiO_2$ (285 nm) wafer and use an atomic force microscope (AFM) to identify unwanted steps and tape residues on the surfaces of the flakes. Using a PC/PDMS stamp, we subsequently pick-up hBN, graphene, hBN, and a graphite flake, which is used as the global bottom gate, and then drop down the whole stack on a $Si/SiO_2$ wafer at elevated temperature. To align hBN and graphene, it is important to precisely determine the crystal axes; under an optical microscope, we visually identify the angles between their sharp atomically defined edges, which corresponds to multiples of 30° reflecting the rotational symmetry of the hexagonal lattice structures [27]. Then, we align graphene and hBN's crystal axis during the stacking process. A narrow (~500 nm in width) top gate is fabricated by a standard e-beam lithography technique, with a proper caution to ensure the adhesion of Ti/Au layer to the hBN top surface.

A couple of devices, with a local top (metal) gate and a global bottom (graphite) gate, were fabricated through a nano-fabrication process on $SiO_2$/p-doped Si wafers, as shown in **Fig. 1(a)** and Figure. S1(b). To improve the contact resistance of our device, we used the Si bottom gate as a contact gate, inducing high charge density near the contacts that are not covered by the bottom graphite. Without the gating for contacts, an unwanted P-N junction may appear at the interface of the graphite and Si gate, disrupting ohmic contacts to the QH edges in the bulk area (Figure. S3). The mobility of the devices is about 70,000 ~ 100,000 $cm^2$ $V^{-1}$ $s^{-1}$ (Figure. S1(c)) and for these graphene heterostructures to exhibit high quality, it was crucial to choose the right contact gate voltage to maintain good contact resistance under high magnetic fields and low temperature.

## 3. Results and discussion

*3.A. Characterization of the bulk properties of the hBN-aligned graphene device*

As shown in **Fig. 1(a)**, we typically measure our multi-terminal devices with the configurations for the longitudinal resistance $R_{xx} = V_{xx}/I$ and the transverse resistance $R_{xy} = V_{xy}/I$ simultaneously, using a lock-in amplifier at the frequency of 17.777Hz and the current bias amplitude of 10 nA. All the measurements were performed at T = 40 mK. In **Fig. 1(b)**, the $R_{xx}$ data measured at $B = 0\ T$ as a function of the top ($V_{tg}$) and bottom ($V_{bg}$) gate voltages show highly-resistive states that appear as multiple straight lines: the lines that appear horizontally in the figure correspond to the global area to become the moiré-induced band insulators, where the density is tuned only by the bottom gate; on the other hand, the density of the local area is tuned by both the local and global gates with the different capacitances, resulting in the insulating phases to appear as the sloped lines [8–10]

To characterize the QH phases in the presence of moiré-induced band gaps, in **Fig. 1(c) and 1(d)**, we performed magnetotransport measurements. We here effectively eliminated the action of the top gate to get the information of the QH phases of the homogeneous bulk of the whole device; this was done by tuning the local top gate voltage so that the density of the local is always the same to that of the global area. In the data, the Chern numbers of the QH states are clearly identifiable with the transverse conductivity $\sigma_{xy} = R_{xy}/(R_{xx}^2 + R_{xy}^2)$ (**Fig. 1(d)**) that shows quantized plateaus at integer multiples of the quantum conductivity, coinciding with the

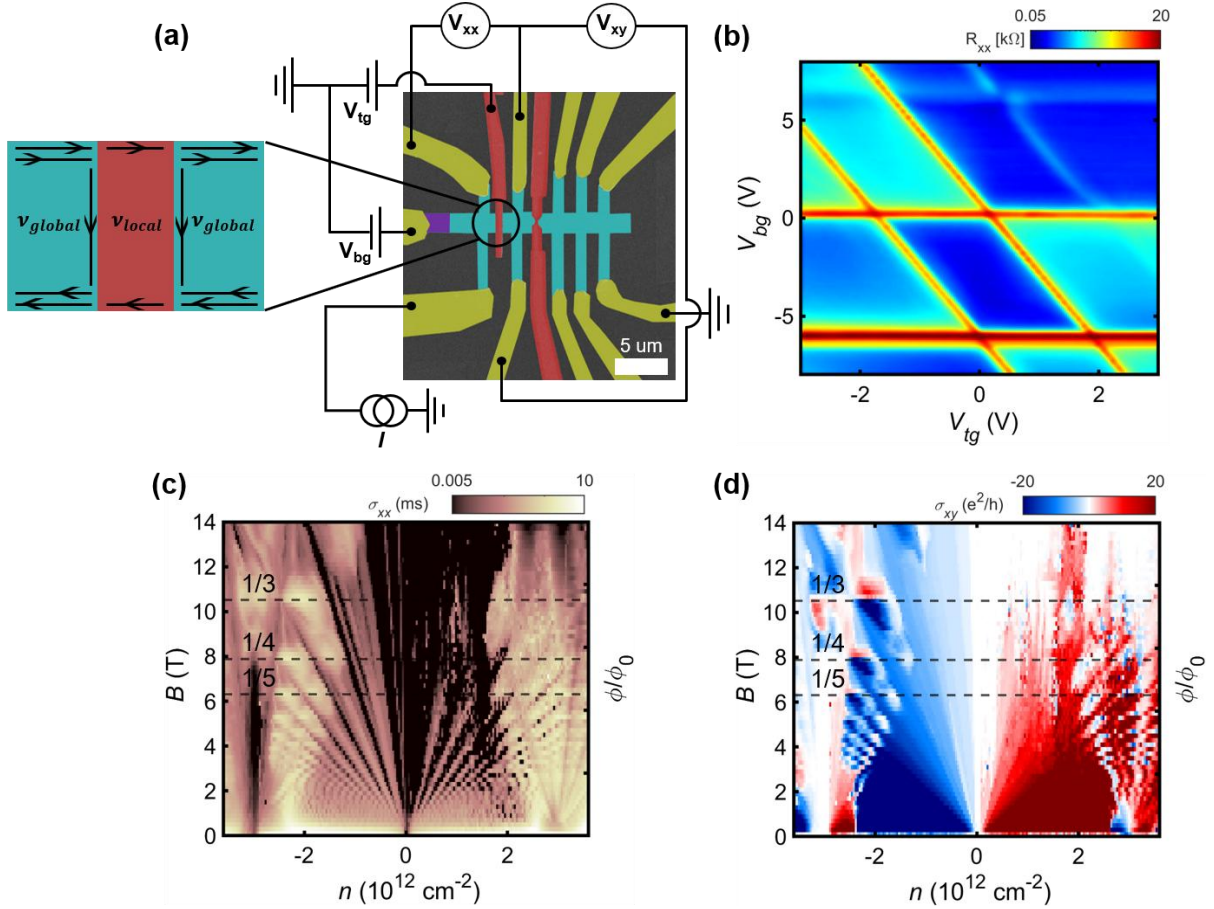

**Figure. 1.** Basic properties of the hBN-aligned graphene with local top gates. (a) A scanning electron microscopy (SEM) image of device A. There are three different gates; one bottom graphite gate and two different top gates. We only analyze the straight top gate, which has 500 nm width. (b) Resistance upon top and bottom gates. Due to the high resistance at hSDP, FDP, and eSDP, nine different lines are observed. The measurement conditions are 40mK and 0T. (c), (d) Landau fan graph with $\sigma_{xx}, \sigma_{xy}$. The $\sigma_{xx}$ becomes zero at QH region and $\sigma_{xy}$ has quantum conductance. The BZ oscillations are observed in the $\sigma_{xx}, \sigma_{xy}$ graphs with black dashed lines.

suppressed longitudinal conductivity $\sigma_{xx} = R_{xx}/(R_{xx}^2 + R_{xy}^2)$ (**Fig. 1(c)**). In these Landau fan graphs, the QH states originating from the hSDP are very noticeable on the hole-doped side. In addition, the horizontal lines (black dashed) are observed at 10.5T ($\phi_0/3$), 7.9T ($\phi_0/4$), 6.3T ($\phi_0/5$) and so on. This is when the lattice periodicity in hBN-aligned graphene becomes commensurate with the cyclotron orbits under a certain magnetic field, leading to the emergence of Brown-Zak (BZ) quasiparticles [23–25], experiencing an effective zero magnetic field at $\phi = \phi_0\, p/q$, where $\phi$ and $\phi_0$ are magnetic flux and flux quantum, and p and q are integer numbers. Based on the sequence, the superlattice parameter is estimated to be 12 nm, corresponding to the twisting angle of 0.6° between the graphene and hBN

*3.B. Conventional PN junction in moiré graphene: p-insulator-n junction*

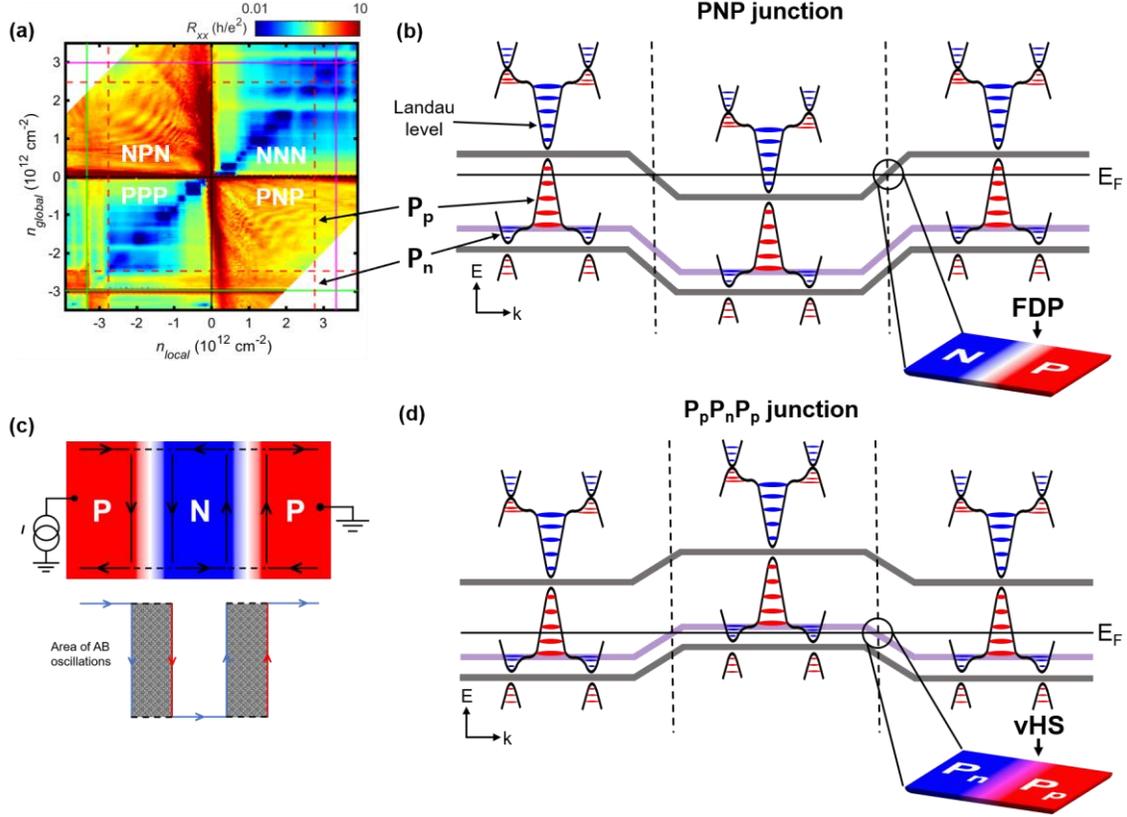

**Figure 2.** The quantum resistance modification upon doping. (a) The $R_{xx}$ according to the charge densities of local and global area. The black, green, magenta solid lines are the FDP, hSDP, eSDP at local and global areas. The red dashed lines represent the emergence of the vHS. (b) Band diagram of PNP junction. The gray strip represents the FDP, crossing the Fermi level. (c) The schematic of Aharonov-Bohm oscillations in PNP junction. The interference loop is formed due to the scattering near the sharp corners. (d) Band diagram of $P_pP_nP_p$ junction. The violet band represents the van Hove Singularity (vHS), crossing the fermi level.

We now investigate the junctions defined by the local gate electrode by setting different charge densities in the global and local-gate areas, denoted as $n_{global}$ and $n_{local}$, respectively. In **Fig. 2(a)**, the longitudinal resistance, measured across the local-gate area, at B = 7.2 T is shown. (Note that the gaps at hSDP and FDP are both visible). First of all, when the junction is tuned to P-N-P and N-P-N doping configurations, i.e., in the bipolar dopings, clear oscillations in magnetoresistance are observed. These oscillations are straightforwardly understood if one considers, for example, a P-N-P junction as formed by two identical P-N junctions, as shown in **Fig. 2(c)**. In a P-N junction formed by a gate-defined smooth chemical potential variation, the interaction and scatterings between the co-propagating QH edge channels at the bipolar doping interface are suppressed due to the presence of the insulating area, separating the channels, where mobile charge carriers are depleted. The effective isolation between the channels doesn't exist at the sharp corners, near the intersections of the sample's physical boundary and the gate-defined interface, partially scattering quasiparticles in-between the channels and therefore acting as beam splitters. Then the insulating area multiplied by the magnetic field plays the role of an interference loop, giving rise to the Aharonov-Bohm (AB) interference patterns.

This phenomenon is qualitatively the same as the one observed in monolayer graphene [12,15]. In hBN-aligned graphene devices, however, the presence of the energy gap at FDP appears to enhance visibility of the AB oscillations, consistently observed across all magnetic field strengths. The insulating phase at FDP (or CNP; $\nu = 0$) is very robust throughout all magnetic fields (see **Fig. 1(c)**). The bulk energy gap is estimated to be ~ 30 meV at B = 0 and further increases due to the enhanced Coulomb interaction in high B fields. This stable energy gap prevents disruption of electron wave coherence, which is essential for clear quantum interference and the observation of the AB effect (see Supplementary Section II).

In our data, the periodicity of AB oscillations is extracted by analyzing the $R_{xx}$ under both fixed and various magnetic fields as the AB oscillations can be described phenomenologically: $G \sim cos(2\pi\Phi/\Phi_0)$, where $\Phi_0 = h/e$ is the magnetic flux quantum and $\Phi = BA$ is the enclosed magnetic flux. At a fixed magnetic field, the AB oscillation area varies with charge density (Figure S4(b)). As the charge densities in the P and N regions increase, the spatial profile of a charge density across the junction narrows, reducing the AB oscillation area of the P-N junction [14–16]. We simulate this oscillation following the model of a previous study in the supplementary material [15], obtaining an AB oscillation area of approximately ~$10^4$ nm$^2$ at 7.2 T. We also evaluate the oscillation visibility through the magnetic field sweeps (Figure S5(c)), obtaining values in the range of 40~60%. This visibility is comparable to previous studies under high magnetic fields [15,16], where the spin and valley degeneracies are lifted. Remarkably, in our moiré graphene device, AB oscillations are still observed in very low magnetic fields (~1T), owing to the moiré-induced FDP gap at the interface of PN junction (Figure S6).

The AB oscillation is a direct consequence of the energy gap at CNP (or FDP in a moiré graphene) and the insulating depletion area formed at a bipolar junction. The inevitable appearance of the large depletion area with an energy gap at a P-N interface is easily explained in an schematic energy diagram of the junction, in **Fig. 2(b)**, where the spatially varying profiles of the bands and the Fermi levels across a junction are visualized. Because the signs of the doping are inverted at the interface, the Fermi level must encounter an energy gap resulting in a strip of an insulating area. In particular, for a smooth interface junction, the area becomes sufficiently large so that the edge channels on either side are well separated and effectively isolated without any significant effect of interaction. Even with gapless monolayer graphene, an application of magnetic field induces a Coulomb-interaction-driven energy gap in high-quality devices, and thus the propagating QH channels in the dipolar junctions are always strongly isolated.

*3.C. A new kind of PN junction: p-vHS-n junction*

In moiré-graphene systems, however, a distinct type of bipolar junction with a unique doping profile can be realized, as depicted in **Fig. 2(d)** - a junction formed by $P_p$ and $P_n$, where the Fermi level resides in the same valence band throughout the junction. Instead of an energy gap, a vHS passes the Fermi level at the junction interface in this case. The density corresponding to the Fermi level passing the vHS is identifiable by locating the density value where the sign of $R_{xy}$ changes in the low field limit and indicated by the red dashed lines in **Fig. 2(a),** for local and global densities. It is important to note that the existence of vHS, and its logarithmic divergence of the density of states (DOS), in an isolated band is protected by the two dimensionality of the system [28]. In contrast to the conventional P-N junction, the $P_p$-$P_n$ junction does not have depletion of charge carriers at the interface, but a strip of dense carriers is expected as the Fermi level passes vHS. This $P_p$-$P_n$ junction, a new type of P-N junction in which the co-propagating QH channels can interact without the insulating gap, is the consequence of both the moiré pattern (via the hSDP) and intrinsic band dispersion of graphene (via the FDP). It would allow us to investigate interactions between co-propagating edges channels at a junction formed at the interface of two systems of opposite chirality, previously inaccessible regime in a high-quality graphene under high magnetic fields; i.e., the full equilibration (FE).

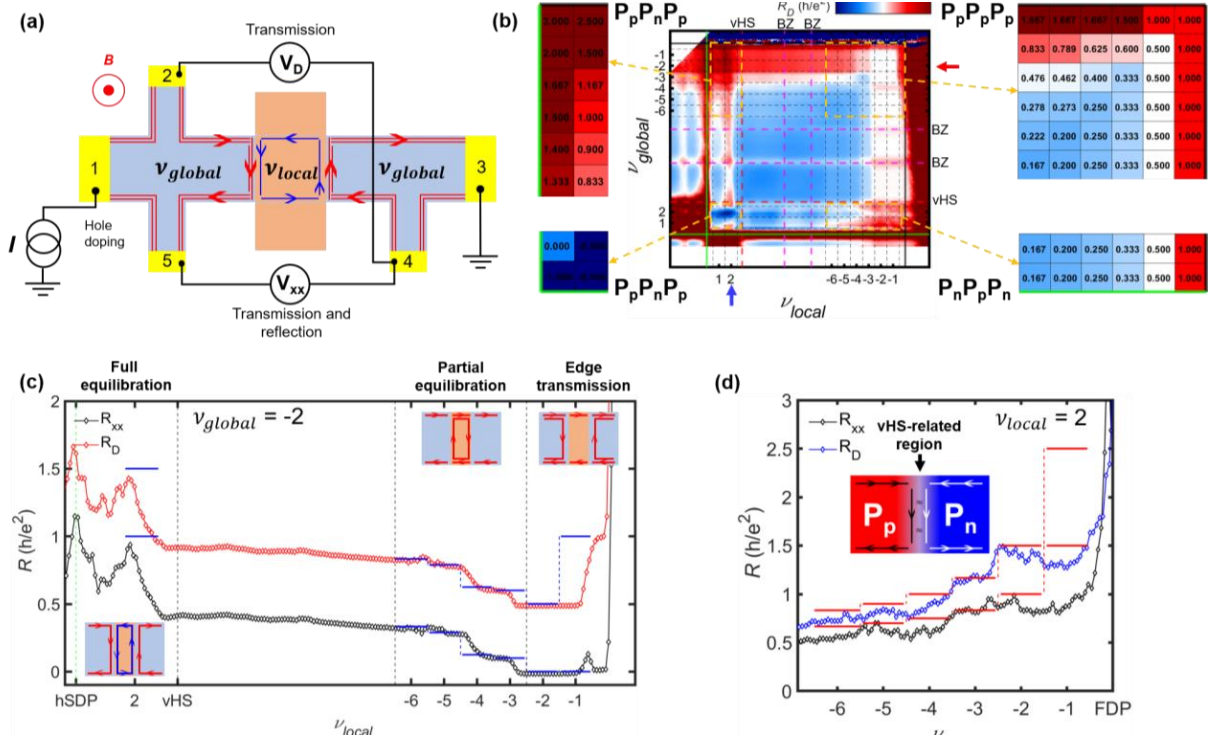

**Figure 3.** Interactions between hSDP- and FDP-mediated QH edges (a) The schematic of transmission and reflections QH edges upon configuration. (b) The modification of $R_D$ according to local and global area's filling factors. The $R_D$ is analyzed under 7.2 T, 40 mK at the PPP doping region. The black dashed lines represent the boundary of the filling factors from the FDP and hSDP. The black and green solid lines are the FDP and hSDP at local and global areas. The magenta and red dashed lines represent the emergence of the BZ quasiparticles and vHS. The red and blue arrows are the line cut positions, $\nu_{global} = -2$ and $\nu_{local} = 2$. The calculation results, utilizing the Landauer-Buttiker formalism, are positioned near the yellow rectangular dashed lines. Note that, beyond the filling factors of $\nu = -6$, the quantisations originating from chiral QH edges become less discernible and, at certain fillings, the influence of BZ quasiparticles becomes dominant, as indicated as dashed magenta lines. (c) Quantum resistance upon configurations with the fixed $\nu_{global} = -2$. The red and black markers represent the $R_D$ and $R_{xx}$. Three different regions are defined by the QH interactions; ET, PE, and FE. Blue lines are the theoretical quantum resistance with Landauer-Buttiker. (d) Quantum resistance upon configurations with the fixed $\nu_{local} = 2$. The blue and black markers represent the $R_D$ and $R_{xx}$, revealing the interaction of QH edges within $P_pP_nP_p$. Red lines are calculation results of $P_pP_nP_p$ junction upon $\nu_{local}$ and $\nu_{global}$. The inset schematic is the $P_pP_n$ junction with vHS-related region.

In our experiment, $P_p$-$P_n$ junctions between QH edges are realized in the moiré-induced isolated band within the $P_pP_pP_p$ region (see **Fig. 2(a)**), where the moiré-induced isolated valence band qualitatively modifies the conventional QH phenomenology of the intrinsic graphene. The diagonal ($R_D$) and longitudinal ($R_{xx}$) resistances, measured across the local-gate area, are used to analyze QH edges interactions in a device tuned to $P_pP_nP_p$ dopings, as shown in **Fig. 3(a)**. The choice of $R_D$ and $R_{xx}$ measurement configurations allow us to analyze quantum resistance and extract the information of the interactions and mixings of QH edge channels, without complicated effects of contact resistance [9,12] (see also Fig. S7). In **Fig. 3(b)**, we plotted $R_D$ of the device A as a function of filling factors in top-gated ($\nu_{local}$) and global areas ($\nu_{global}$), where the filling factors for compressible transitions between incompressible QH states and for the region of the enhanced effect of BZ quasiparticles are all accounted for and indicated by black and magenta dashed lines. Importantly, four distinct filling-factor regions, separated by the lines for vHS (red dashed), are identifiable. (The additional measurements using different configurations, such as $R_{53}$ and $R_{52}$, and data of device B can be found in Fig. S8 and S9 of the supplementary materials).

For quantitative comparison with theoretical models, in **Fig. 3(b)**, the Landauer-Buttiker formalism was used to systematically evaluate the expected quantum resistances by using various possible edge-equilibration models (such as ET, PE and FE etc.; see subset schematics of the models in **Fig. 3(c)**). The results that best-match the experimental data are presented in the panels outside. The assignment of filling factors for the calculation is based on the Chern number measured in the bulk experiment in Section 3.A. While evaluating the topology of the Hofstadter bands and assigning the Chern numbers need a full quantum mechanical treatment in principle, a semi-classical picture is effective in the presence of large energy gaps at hSDP and FDP, with which the Chern numbers can be interpreted as positive (or negative filling factors) of LLs originating from the band minimum (or maximum), we thus assigned negative filling factors to the states energetically lying above vHS and positive filling factors to the states below vHS, referenced from the adjacent band gaps, FDP and hSDP, respectively. Based on this semi-classical filling factor assignment, we classify the doping configurations into four distinct regions: $P_pP_nP_p$, $P_pP_pP_p$, $P_nP_pP_n$, and $P_nP_nP_n$ (see **Fig. 3(b)**). A comparison between the calculated quantum resistances based on the FE model and the experimental data in **Fig. 3(b)** shows *quantitatively good* agreement in the $P_pP_nP_p$ and $P_pP_pP_p$ regions. On the other hand, significant deviations from the theoretical models are observed in the $P_nP_pP_n$ and $P_nP_nP_n$ regions; we attribute these to disorders likely due to contaminants on the exposed hBN top surface in the global area unlike the local-gate area which is encapsulated by the top and bottom gate graphite. It is known that the energy gap at hSDP is highly susceptible to disorders [29].

More in-depth comparison of the measured quantum resistances to theoretical equilibration models in the $P_pP_nP_p$ and $P_pP_pP_p$ regions of the moiré-induced isolated band are discussed in the following. The line cut at $\nu_{global} = -2$ (indicated by the red arrow in **Fig. 3(b)**) is plotted in **Fig. 3(c)**. With spin and valley degrees of freedom in an orbital LL, the filling factor of 2 corresponds to the orbital gap between ZLL (N = 0) and the first excited LL (N = 1). In **Fig. 3(c)**, the measured quantum resistance values match fairly well with the expected values evaluated based on one of the equilibration models, indicated by blue solid lines (also see the Supplementary Materials for the model analysis and the discussion of the fractional quantum resistance in the $P_pP_pP_p$ region, as summarised in the Table S1.) Notably, at $\nu_{local} = 2$ and $\nu_{global} = -2$ where the sample is in the $P_pP_nP_p$ state, the quantum resistances $R_{xx}$ and $R_D$ have the values that are only consistent with the FE model.

Now, in **Fig. 3(d)**, we measured the line cut at $\nu_{local} = 2$ (indicated by the blue arrow in **Fig. 3(b)**), where the local-gate area is kept in a $P_n$-doped state for analysis. The measured quantum resistances closely follow the calculated results based on the FE model across a wide range of $\nu_{global}$, indicated by the red ($R_D$ and $R_{xx}$) solid lines. Both $R_D$ and $R_{xx}$ remain close to the theoretical values in $\nu_{global}$= -2, -3, -5, and -6 as shown in **Fig. 3(d)** and **Table 1**, while some poor agreement is noticeable in $\nu_{global}$= -1 and -4 (**Fig. 3(d) and Table 1**), likely due to insufficient energy gaps between LL subbands and consequential under-developed edge channels. Overall, the consistent observations of the $P_pP_nP_p$ configuration at different conditions and temperatures (Figure. S11 and S12) further supports the emergence of a new type of PN junction without the depletion region, giving rise to full interaction between the co-propagating channels of opposite chirality.

| $\nu_{global}$ | -1 | -2 | -3 | -4 | -5 | -6 |
|---|---|---|---|---|---|---|
| $R_D$ (exp) | 1.330±0.06 | 1.409±0.04 | 1.162±0.03 | 0.870±0.06 | 0.810±0.03 | 0.739±0.03 |
| $R_D$ (cal) | 2.500 | 1.500 | 1.167 | 1.000 | 0.900 | 0.833 |
| $R_{xx}$ (exp) | 0.893±0.06 | 0.872±0.06 | 0.853±0.05 | 0.629±0.04 | 0.627±0.04 | 0.553±0.02 |
| $R_{xx}$ (cal) | 1.500 | 1.000 | 0.833 | 0.750 | 0.700 | 0.667 |

**Table 1**. The quantum resistances upon filling factors within fixed $\nu_{local}$= 2. The quantum resistances regarding measurement configurations are represented with the $h/e^2$ unit. The calculation and experimental data show the consistent values at $\nu_{global}$= -2, -3, -5, and -6.

*3.D. Possible mechanism of the enhanced equilibrium between co-propagating channels: vHS-induced metallic phase in QH limit*

The occurrence of the full equilibrium resembles the equilibration previously observed in *rough* PN junctions, in which the roughness in the interfaces promotes scattering events between otherwise *isolated* co-propagating edge channels. In our devices, due to their high quality, scatterings between channels are nearly suppressed as a result of expectable intentional sharp corners near the sample's physical edges with a good isolation between the co-propagating channels in a straight section of the interface, giving rise to the AB oscillations as seen in **Fig. 2(a)**. The near complete equilibration observed in $P_pP_n$ junctions therefore requires a unique mechanism to be explained: we attribute this to the inevitable existence of an area of a phase based on the phenomena of magnetic breakdown near vHS.

When a junction has a $P_pP_n$ interface, there must be an area in the form of a strip, where the local Fermi level is near vHS, in the middle, as shown in the inset schematic of **Fig. 3(d)** (see also **Fig. 2(d)**). In this area, the unique band dispersion near the vHS is known to lead to multiple electron-like and hole-like Fermi pockets in the momentum space [30–32]. The magnetic breakdown occurs when the orbits move closer together in momentum space and the tunnelings between the orbits become highly proliferated [28,33]. In this state, the filling factors start to lose their meaning and one expects that electron- and hole-like excitations are present simultaneously, without distinct net chirality as a whole, contributing and modifying the overall transport even in the QH limit. We argue that this strange metal-like state in a high magnetic field induced by magnetic breakdown is potent to greatly promote interaction and equilibration across the $P_pP_n$ interface and lead to FE.

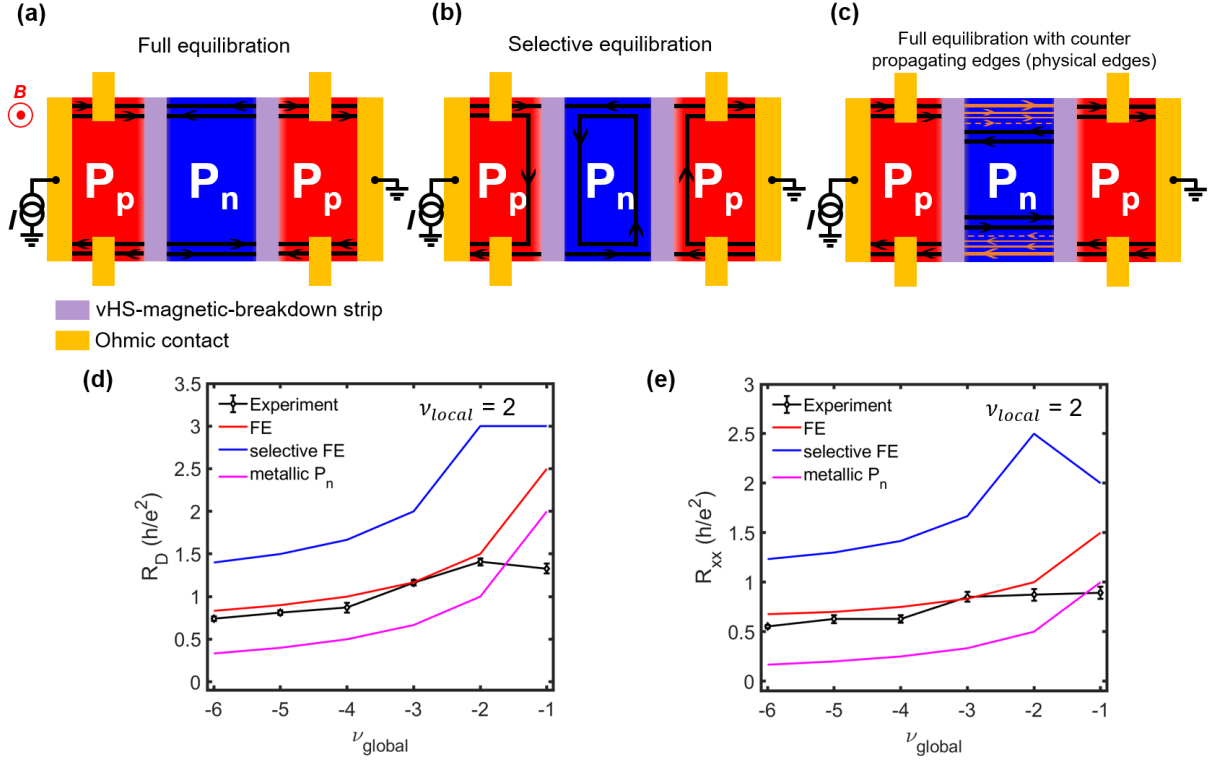

**Fig. 4** Comparison of three different models; full equilibration, selective equilibration, and counter propagating edges in $P_n$ (a) Schematic of the Full equilibration. The violet area represents the equilibration region, which is equivalent to the vHS-magnetic-breakdown strip. (b) Schematic of the selective equilibration. Retained QH edge channels emerge in both local and global areas. (c) Schematic of the counter propagating edges in $P_n$. Counter propagating edges are preserved at the device edges as the opposite chirality is observed after vHS. (d), (e) The resistances of calculated and experiment value of the $R_{xx}$ and $R_D$. The experimental value is close to the full equilibration model.

To corroborate this picture, in **Fig. 4(a-c)**, three models based on the enhanced equilibration of the 'vHS-magnetic-breakdown' strip (thin violet areas) are considered as possible scenarios to explain the measurement; namely, full equilibration, selective equilibration, and full equilibration with counter propagating edges. We first model that each of $P_n$- and $P_p$-doped area has chiral channels that correspond to a semi-classical filling factor we assigned previously, and that the area of magnetic breakdown near the vHS behaves as an interface that facilitates interaction as shown in **Fig. 4(a)** and **(b)**. In the first scenario, the strip fully equilibrates the copropagating channels. In the second scenario, only a portion of the QH edge channel is equilibrated near the strip, thus which is called selective equilibration, as shown in **Fig. 4(b)**. This case is expected to display higher resistance than the full equilibration case, as these configurations represent intermediate states between FE and fully-insulating states among QH channels. However, in the third scenario, as shown in **Fig. 4(c)**, we model an edge structure of many counter-propagating channels along the physical boundary in the local-gate area, to account for the possibility that the number of channels is actually different from the filling factor we semi-classically assigned for the $P_n$ doping. For example, complicated Hofstadter bands of various Chern numbers potentially induce many counter propagating edges in the $P_n$ region. Note that in this case, the whole local-gate area participates in equilibration between the global areas, effectively forming a $P_p$-metal-$P_p$ junction. A comparison of the experimental values to the calculated values of the three scenarios based on the Landauer-Buttiker formalism is shown in **Fig. 4(d)** and **(e)**.

Overall, the experimental data are well explained by the full equilibration model, suggesting that the strip formed by the unique band dispersion near vHS strongly enhances the interaction between QH channels. We emphasize that this $P_pP_n$ junction, a new type of PN junction, is a unique realization of co-propagating QH channels, in which the interactions between P and N doping QH channels arise due to the presence of the strange vHS-induced metallic area instead of a depletion area as in a conventional PN junction. A future theoretical work to better describe the strange phase may need a full quantum mechanism treatment beyond any semi-classical picture.

## 4. Conclusion

In conclusion, we investigated the interactions between moiré-modified QH edges by utilising the local and global gates in a hBN-aligned graphene device. Because of this tunability, two different kinds of PN junctions, one with an insulating gap and the other with a strip with enhanced equilibration due to the singular DOS near vHS, at the interface are realized, allowing comparison of the two limiting cases of equilibration between copropagating channels in a bipolar junction. We observed the edge transmission, partial equilibration, full equilibration, and AB oscillations in the single device due to the unique band structure of a moiré graphene. The full equilibration observed in the $P_pP_nP_p$ configuration offers deeper insights into the interplay between the QH edge channels in a moiré potential. Our research provides a platform to study the interactions between moiré-induced QH edge channels and suggests potential applications in moiré-based heterostructures as future quantum information devices [23,34–37].

## Supplementary material

See the supplementary material for additional information about the fabrication process and characterize the device properties.

## Acknowledgments

This work was supported by the National Research Foundation of Korea grants funded by the Ministry of Science and ICT (Grant Nos. RS-2025-23525425, RS-2020-NR049536, and RS-2023-00258359), SNU Core Center for Physical Property Measurements at Extreme Physical Conditions (Grant No. 2021R1A6C101B418), and Creative-Pioneering Researcher Program through Seoul National University. J.J. K.W. and T.T. acknowledge support from the JSPS KAKENHI (Grant Numbers 21H05233 and 23H02052) and World Premier International Research Center Initiative (WPI), MEXT, Japan.

## Author contributions

W.B.C. and M.J. performed the experiments. W.B.C. and J.J. analyzed the data. K.W. and T.T. provided hBN crystals. J.J. supervised the project. W.B.C. and J.J. wrote the manuscript with input from all authors.

## Author declarations

### Conflict of interest

The authors declare no competing interests.

## Data availability

The data that support the findings of this study are available from the corresponding authors upon reasonable request.

# Supplementary Information for "Interaction of moiré-induced quantum Hall channels in a locally gated graphene junction"

Won Beom Choi[1,2], Myungjin Jeon[1,2], K. Watanabe[3], T. Taniguchi[4], Joonho Jang[1,2]*

[1] Department of Physics and Astronomy, and Institute of Applied Physics, Seoul National University, Seoul 08826, Korea

[2] Center for Correlated Electron Systems, Institute for Basic Science, Seoul 08826, Korea

[3] Research Center for Electronic and Optical Materials, National Institute for Materials Science, 1-1 Namiki, Tsukuba 305-0044, Japan

[4] Research Center for Materials Nanoarchitectonics, National Institute for Materials Science, 1-1 Namiki, Tsukuba 305-0044, Japan

*Corresponding author's e-mail: joonho.jang@snu.ac.kr

## I. Basic sample information

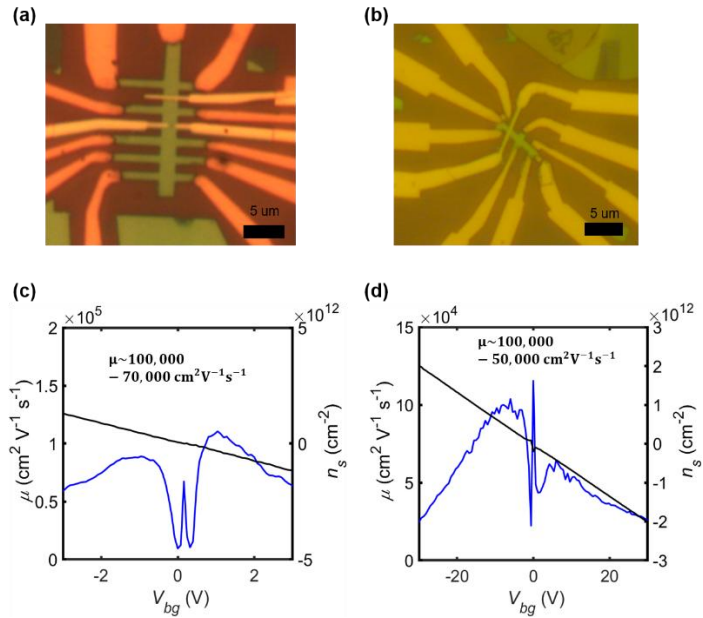

**Fig. S1** (a), (b) Optical image of the device A and B. (a) is the picture of device A with two top gates. (b) is the picture of device B with one top gate. (c), (d) The mobility of the device A and B. The device A and B show the 50,000 - 100, 000 cm² V⁻¹ s⁻¹ upon bottom gate voltage.

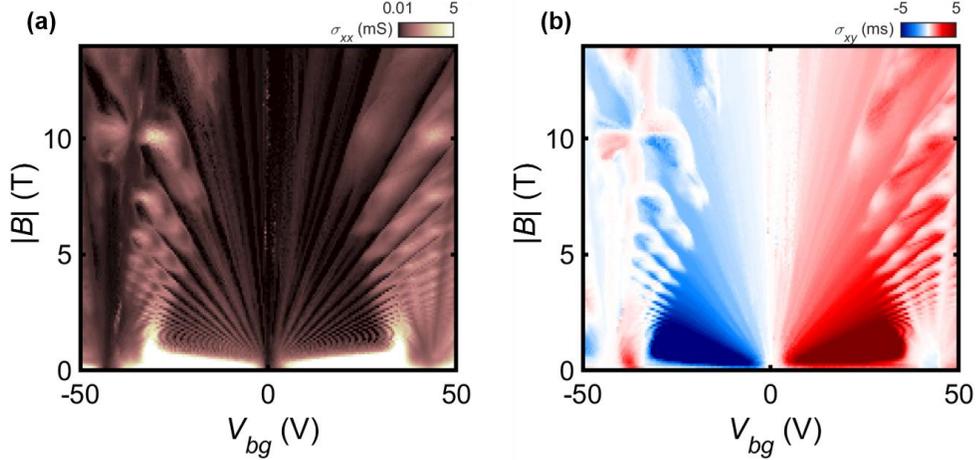

**Fig. S2** (a), (b) Landau fan graph of the device B with $\sigma_{xx}$ and $\sigma_{xy}$. The hSDP mediated QH effects are observed in $\sigma_{xx}$ and $\sigma_{xy}$.

The Hall mobility of device A and B was measured by varying the magnetic field up to 1T. Mobilities in the range of 50,000 to 100,000 cm² V⁻¹ s⁻¹ were obtained as a function of carrier doping, as shown in **Fig. S1**. These values may be underestimated due to the FDP in hBN-aligned graphene [17–19]. Device A incorporates a graphite bottom gate, whereas device B utilizes a silicon bottom gate. We also estimate our device quality with a Landau fan graph as shown in **Fig. S2,** where complete lifting of spin and valley degeneracy is observed above 6T. All measurements were conducted at a base temperature of 40 mK. These results collectively demonstrate that devices A and B exhibit high-quality characteristics of hBN-aligned graphene.

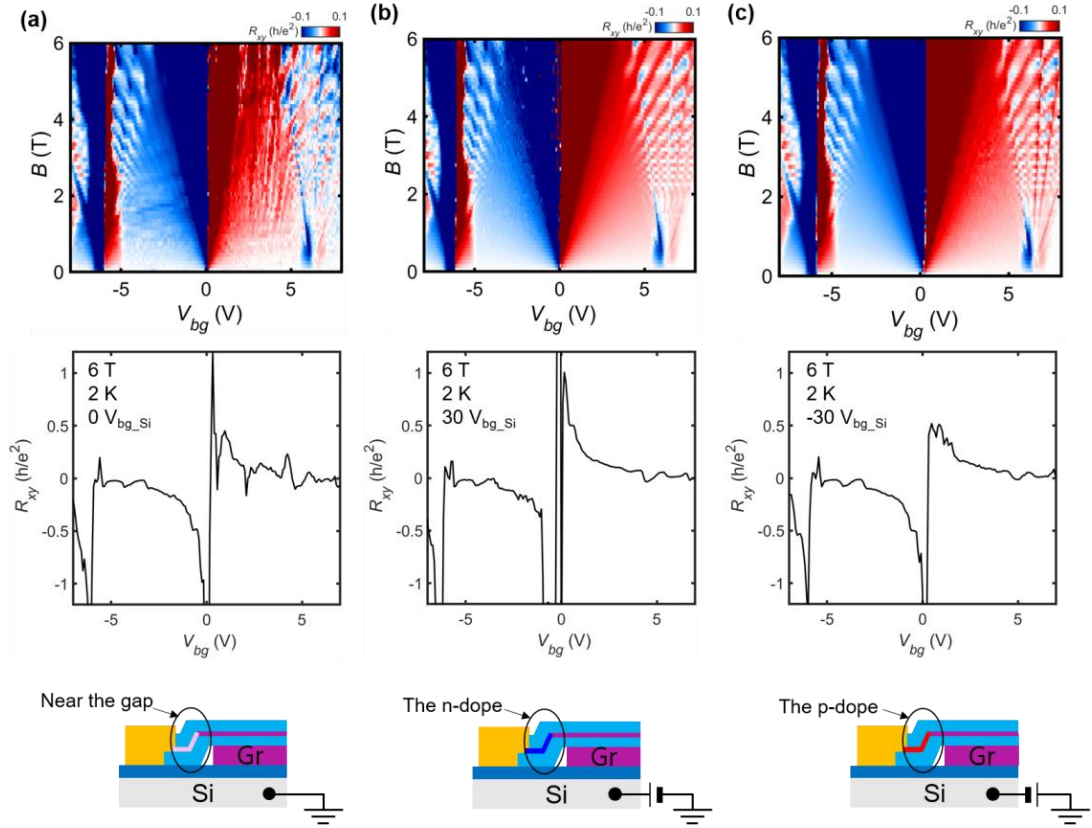

**Fig. S3** (a), (b), (c) The Landau fan graph upon Si doping at device A. The Si only tuning region exists due to the sample geometry. The Si gate is tuned at ground, plus voltage and minus voltage at (a), (b), and (c). These types of doping represent the gap, n-doping, and p-doping at electrode regions.

In device A, the Si and graphite gates can be independently controlled. The main region is tuned with the graphite whereas the parts of the electrode regions are selectively gated with the Si gate to prevent current leakage from the graphite gate to the electrodes. As shown in **Fig. S3**, the QH resistances are modulated by the bottom gate voltage, as the formation of a depletion region between main region and electrode region disturbs the charge path. Especially, the QH resistance becomes blurred when the Si gate is set as the ground, bringing the electrode regions close to the band gap (FDP). In contrast, the quantum resistances are clear when both the Si and graphite gates induce the equivalent doping types, as shown in **Fig. S3(b) and (c)**. The bottom Si gate was fixed at -15 $V_{bg}$ to focus on the p-doping regime.

## II. AB oscillations in hBN-aligned graphene

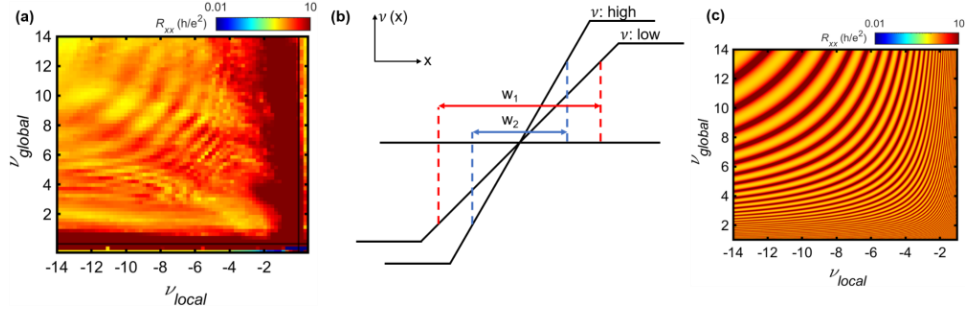

**Fig. S4** (a) The measured resistance in the NPN region. The resistance oscillations are observed according to local and global filling factors. (b) filling factor modification in the interface of the local and global area. An effective area W is modified according to filling factors (or doping levels). (c) Simulated resistance in the NPN region. The parabolic oscillations correspond to the experiment data.

We now analyze the AB oscillations in the NPN (or PNP) region. The variation of $R_{xx}$ with filling factors is shown in **Fig. S4(a)**, where a parabolic dependence dominates at high filling factors in both the local and global areas. The schematic in **Fig. S4(b)** represents the interface between P- and N-doping areas (or global and local areas). The effective width of the PN junction, denoted as $W_1$ and $W_2$, depends on the respective filling factors. The difference in filling factors across the PN interface defines the interfacial slope, as shown in **Fig. S4(b)**. This slope increases (or decreases) as the filling factors in the global and local regions become higher (or lower), corresponding to stronger (or weaker) doping levels. Therefore, variations in local and global filling factors modulate the width of the insulating area at the PN interface (see **Fig. 2(c)** in the main text), leading to AB oscillations through local and global doping control. We simulate these AB oscillations at a fixed magnetic field by modeling the schematic in **Fig. S4(b)** using hyperbolic tangent function [15]. The resulting simulation, shown in Fig. S4(c), reproduces the experimental oscillation behavior with tuning parameters corresponding to a magnetic field of 7.2 T and an effective insulating area of approximately $S \sim 10^4$ nm$^2$.

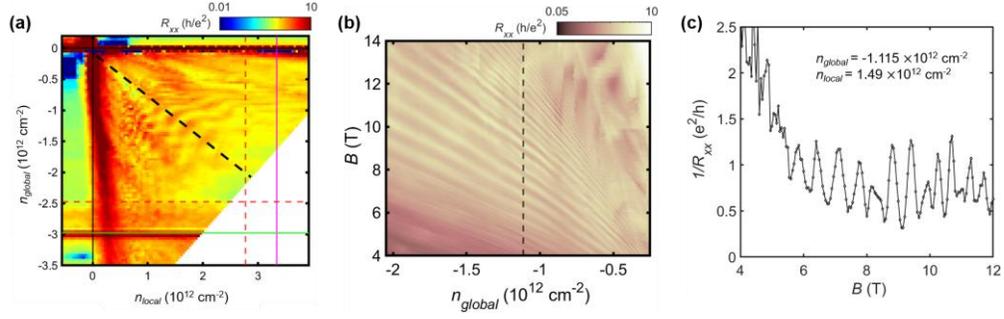

**Fig. S5** (a) The resistance graph with the local and global area's charge density in the PNP region under 7.2T and 40 mK. We perform the line cut following the black dashed line. (b) Field sweep of the black dashed line in (a). (c) Line cut of (b). We fixed the local and global area's charge density at $1.49 \times 10^{12}$ and $-1.115 \times 10^{12}$ cm$^{-2}$ respectively.

The effective AB oscillation area (or insulating area) is also determined by varying the magnetic fields, as the AB oscillations follow the relation $g \propto cos\left(2\pi \frac{BA}{\phi_0}\right)$, where g is the conductance, B is the magnetic field, A is the AB oscillation area, and $\phi_0$ is the magnetic flux quantum. The magnetic field sweeps are performed in the PNP region according to black dashed line in **Fig. S5(a)**. The AB oscillations become apparent with increasing magnetic field and charge density, as shown in **Fig. S5(b)**. The oscillation period obtained from the magnetic field sweep varies with charge density, reflecting changes in the effective AB oscillation area. The effective insulating area is extracted from **Fig. S5(c)**, which displays the sinusoidal oscillations as a function of the magnetic field. The estimated effective area is approximately $6.4 \times 10^3$ nm$^2$, consistent with the simulation value of ~$10^4$ nm$^2$. The visibility of the oscillations in this line cut is approximately 40 ~ 60%.

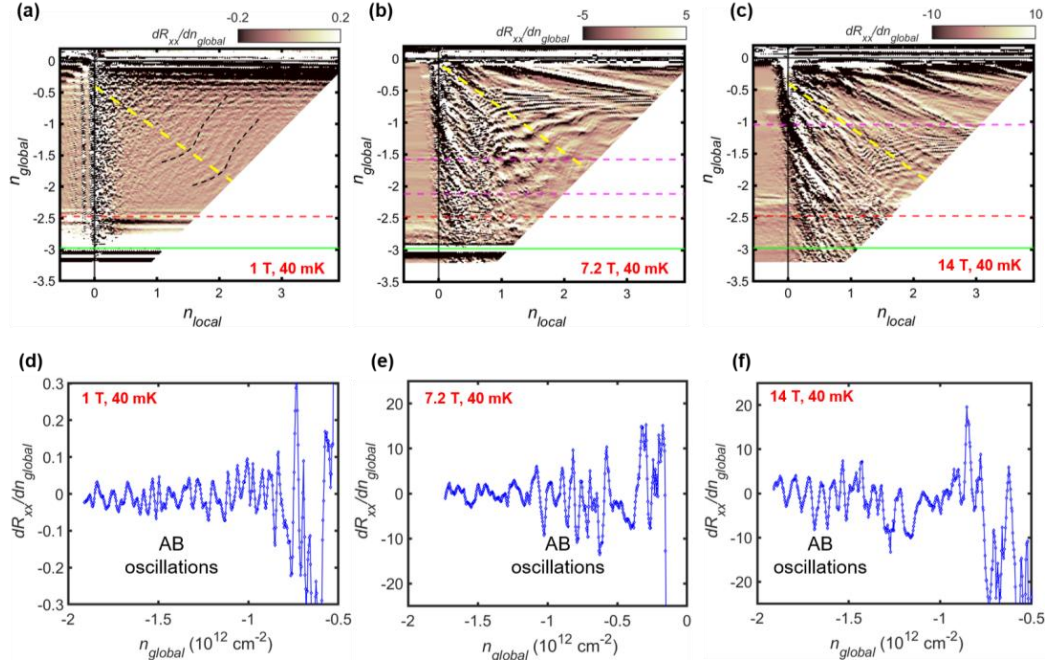

**Fig. S6** (a) - (c) The AB oscillation measured at various magnetic fields. The AB oscillation is observed even in 1T. (d) - (e) Line cut graph at 1T, 7.2T, and 14T. The position of the line cuts corresponds to the yellow dashed line indicated at the PNP junction.

Owing to the moiré-induced FDP gap, AB oscillations are observed over a wide range of magnetic fields. These AB oscillations appear not only at high magnetic fields but also in low magnetic fields (~1 T), as shown in Fig. **S6(a)-(c)**. The observation of AB oscillations at 1T suggest that the presence of moiré-induced FDP gap enhances the formation of the insulating area, as shown in **Fig. S6(d)**.

## III. QH edge equilibration in hBN-aligned graphene

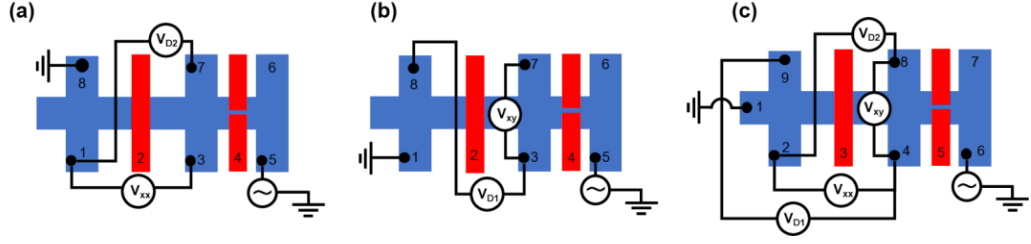

**Fig. S7** Different configurations according to the top gate. (a) and (b) represent the actual measurement configuration. (c) is the effective configuration of the (a) and (b).

The QH edge channels follow the Landauer formalism, considering the nonlocal effects. Therefore, the quantum resistance varies depending on the measurement configuration. This effect is especially dominant at the 2DEG device with local top gate, where different combinations of filling factors lead to distinct regimes such as edge transmission, partial equilibration, and full equilibration. We measured several different configurations like **Fig. S7(a)** and **(b)** in device A and B to observe three different regimes. By adding an additional electrode to the current device, all configurations can be measured simultaneously, as shown in **Fig. S7(c)**. The quantum resistance for each configuration and corresponding filling factor combination is analyzed using the Landauer-Buttiker formalism.

Edge state transmission

$$R_{D1} = \frac{h}{e^2} \cdot \frac{1}{|v_{local}|}, \quad R_{D2} = \frac{h}{e^2} \cdot \left( \frac{|v_{global}| \cdot |v_{local}|}{|v_{global}| - 2|v_{local}|} \right)^{-1},$$

$$R_{xx} = \frac{h}{e^2} \cdot \left( \frac{1}{|v_{local}|} - \frac{1}{|v_{global}|} \right), \quad R_{xy} = \frac{h}{e^2} \cdot \frac{1}{|v_{global}|}$$

Partial equilibration

$$R_{D1} = \frac{h}{e^2} \cdot \left( \sum_{i=\uparrow,\downarrow} \left( \frac{|v_{local,i}| \cdot |v_{global,i}|}{2|v_{local,i}| - |v_{global,i}|} \right) \right)^{-1}, \quad R_{D2} = \frac{h}{e^2} \cdot \left( -\sum_{i=\uparrow,\downarrow} |v_{local,i}| \right)^{-1},$$

$$R_{xx} = \frac{h}{e^2} \cdot \left( \sum_{i=\uparrow,\downarrow} \left( \frac{|v_{local,i}| \cdot |v_{global,i}|}{2|v_{local,i}| - |v_{global,i}|} \right) \right)^{-1} - \frac{h}{e^2} \frac{1}{|v_{global}|}, \quad R_{xy} = \frac{h}{e^2} \cdot \frac{1}{|v_{global}|}$$

Full equilibration

$$R_{D1} = \frac{h}{e^2} \cdot \left(\frac{1}{|\nu_{local}|} + \frac{2}{|\nu_{global}|}\right), R_{D2} = \frac{h}{e^2} \cdot \frac{1}{|\nu_{local}|},$$

$$R_{xx} = \frac{h}{e^2} \cdot \left(\frac{1}{|\nu_{local}|} + \frac{1}{|\nu_{global}|}\right), R_{xy} = \frac{h}{e^2} \cdot \frac{1}{|\nu_{global}|}$$

| $\nu_{local}$ | -1 | -2 | -3 | -4 | -5 | -6 |
|---|---|---|---|---|---|---|
| $R_D$ (exp) | 0.555±0.120 | 0.486±0.001 | 0.593±0.006 | 0.662±0.064 | 0.791±0.017 | 0.823±0.011 |
| $R_D$ (cal) | 1.000 | 0.500 | 0.600 | 0.625 | 0.789 | 0.833 |
| $R_{xx}$ (exp) | 0.022±0.051 | -0.015±0.001 | 0.097±0.004 | 0.171±0.067 | 0.298±0.02 | 0.315±0.013 |
| $R_{xx}$ (cal) | 0.5 | 0 | 0.100 | 0.125 | 0.289 | 0.333 |

**Table S1**. The quantum resistances upon filling factors within fixed $\nu_{global}$=2; P$_p$P$_p$P$_p$ region. The quantum resistances regarding measurement configurations are represented with the h/e$^2$ unit. The calculation and experimental data show the consistent values at $\nu_{local}$= -2, -3, -5, and -6. The experimental quantum resistances in $\nu_{local}$= -1 and -4 are distant from the calculation results. Therefore, plateaus do not appear due to the lack of full degeneracy lift.

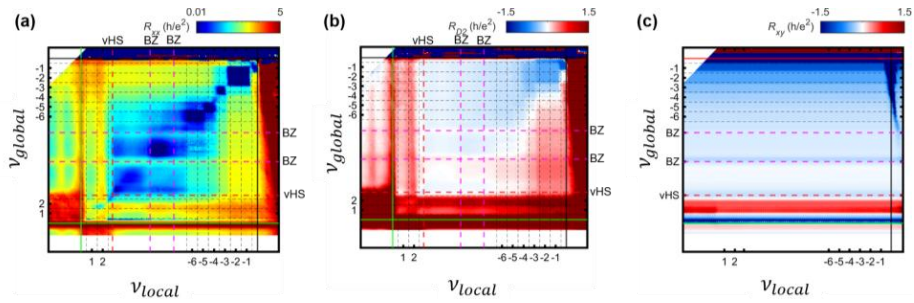

**Fig. S8** The device A's resistance graph under 7.2 T and 40 mK; R$_{xx}$, R$_{D2}$, and R$_{xy}$. (a) and (b) represent the filling factors of local and global regions. (c) represents the global region filling factors only.

Based on these results, we compare the experiment data with the theoretical predictions as shown in **table 1** and **table S1**. The calculation result of the $P_pP_nP_p$ region is driven from the upper formulas and analyzed in the main text. For the $P_pP_pP_p$ region, the quantum resistances are compared with the theoretical values in fixed $v_{global}=2$, as shown **Fig. 3(c)** and **table S1**. Overall, the experimental quantum resistances are consistent with the theoretical predictions, excepting $v_{local}=$ -1 and -4 due to the lack of full degeneracy lift in 7.2 T. Spin polarization is considered in these calculations, matching with the measurement results. These results reveal that the QH edge channels are well developed, and the interactions among the QH edge channels - edge transmission and partial equilibration - are clearly observed across different configurations, as shown **Fig. S8**.

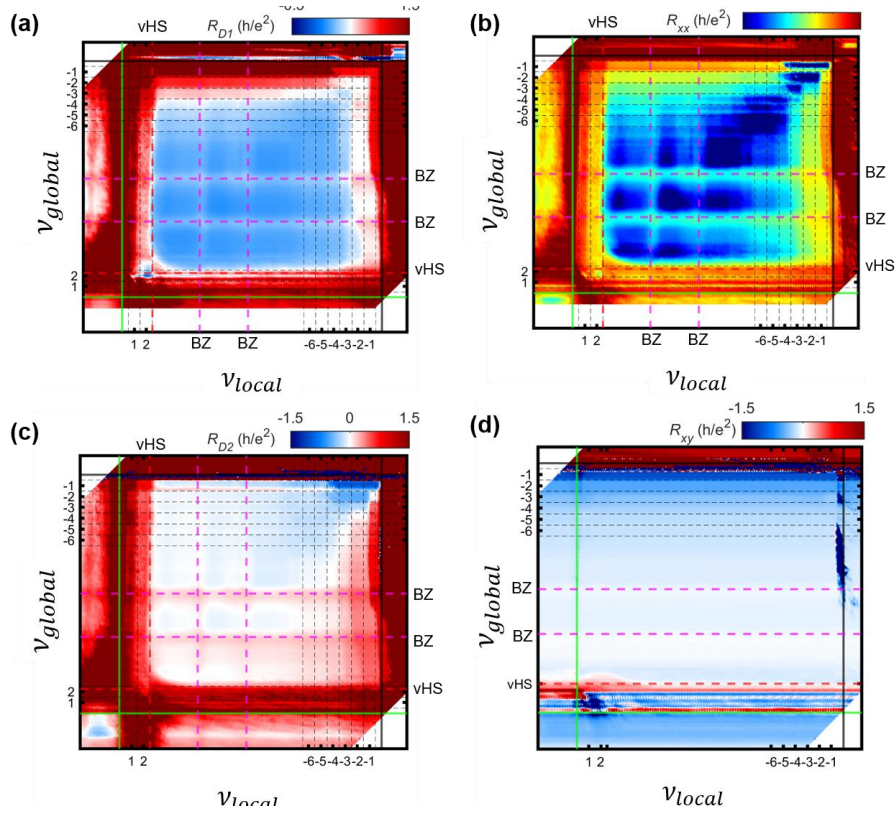

**Fig. S9** The device B resistance graphs upon measurement configurations and charge density in local and global regions. The resistance modifications with all configurations are observed in 5.5 T and 40 mK.

The device B exhibits consistent behavior as shown in **Fig. S9**, especially in $P_pP_nP_p$ region, and displays better-resolved filling factor $|v| = 1$ compared to device A. The device B also shows more clear filling 1 than that of the device A. However, the filling factors becomes less distinct beyond $|\nu| = 1$ in device B due to the low device quality. Despite these limitations, reproducible evidence of the new type of PN junction is obtained, revealing robust full equilibration in hBN-aligned graphene.

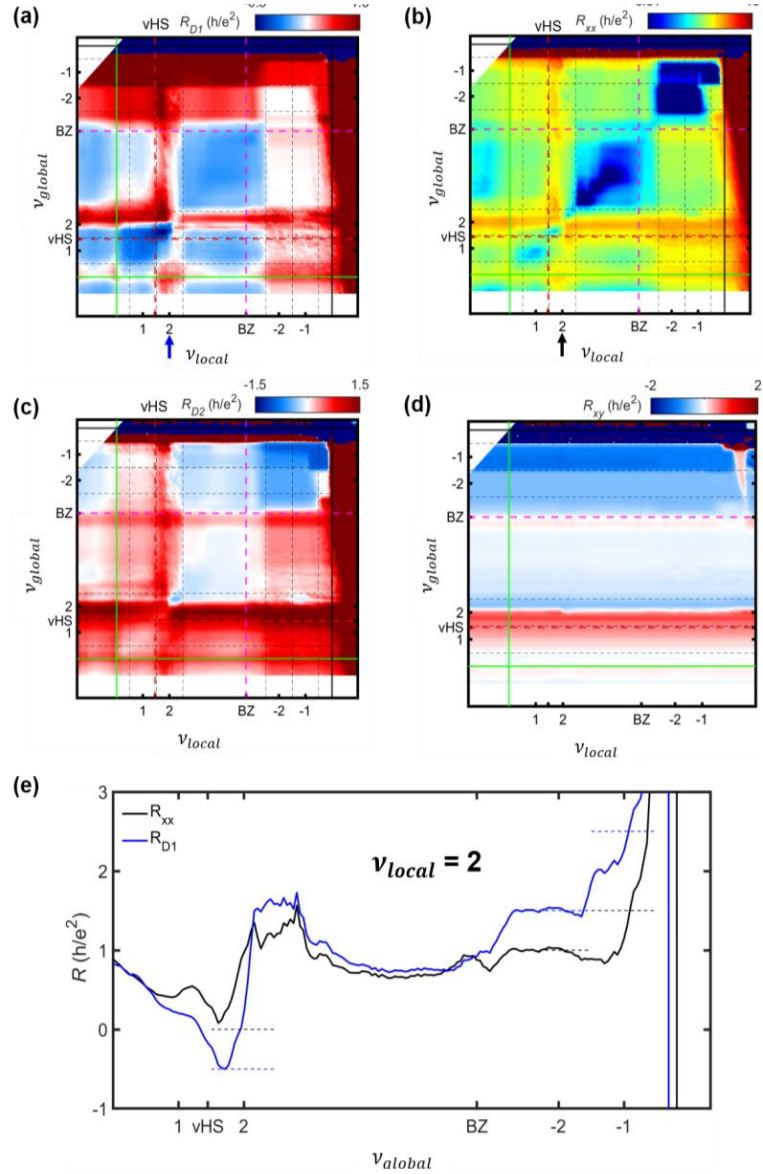

**Fig. S10** The device A resistance graphs upon measurement configurations and charge density in local and global regions. The resistance modifications with all configurations are observed in 14 T and 40 mK. (a), (b), (c), and (d) are $R_{D1}$, $R_{xx}$, $R_{D2}$, and $R_{xy}$ with top and bottom gates. (e) represent the line cut of $\nu_{local} = 2$.

The spin degeneracy lift becomes evident at high magnetic fields due to the Zeeman effect. We measure the device A at the 14T to investigate the QH edge interaction with spin polarization as shown in **Fig. S10**. In this field, the BZ quasiparticles and other Hofstadter butterfly-related signals are dominant (**Fig. S10(d)**), thus only the low integer filling factors are visible; $|\nu| = 1$ and 2. With this condition, the full equilibration -$P_pP_nP_p$ configuration- is observed as shown in **Fig. S10(e)**. The line cut graph in **Fig. S10(e)** shows that the quantum resistance of $P_pP_nP_p$ junction is correspond with the theoretical prediction (black and blue dashed lines) at $\nu_{local} = 2$. The interactions between QH edge channels of opposite doping types are enhanced at high fields due to clear superlattice-mediated QH edge channels. Furthermore, the $P_nP_nP_n$ region exhibits quantum resistance plateaus at -1/2 and 0 $h/e^2$ with $R_{D1}$ and $R_{xx}$, which were not measured under 7.2T. The narrowing of the $\nu_{local} = 2$ may be related to the broadening of the vHS, revealing the counter propagating edges.

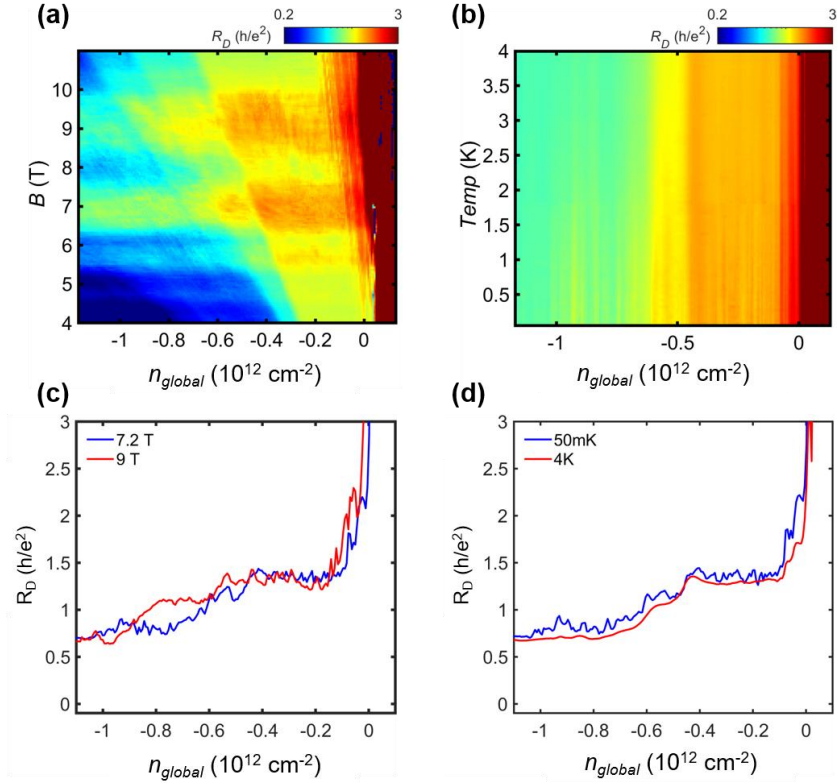

**Fig. S11** The magnetic field and temperature dependence of the $P_pP_nP_p$ region. (a) Resistance as a function of magnetic field and charge density. The charge density is tuned with the fixed $\nu_{local}$ = 2 and varying $\nu_{global}$. (b) Temperature-dependent resistance measured at 7.2 T with $\nu_{local}$ = 2. (c) Line cuts from (a) at 7.2 T and 9 T, showing that the plateaus broaden at 9T. (d) Line cut from (b), where the temperature is varied from 50 mK to 4 K. The fluctuations of plateaus decrease with increasing temperature.

The QH edge channels become more clear at high magnetic fields and low temperatures due to the enhancement of Zeeman energy and suppression of thermal energy. Therefore, we measure the resistance of the $P_pP_nP_p$ region upon magnetic field and temperature, as shown in **Fig. S11**. The QH regions increase upon the magnetic field, originating the wide plateaus as shown in **Fig. S11(c)**. A blurred $P_pP_nP_p$ region is observed between 7 and 9T, corresponding to the presence of the BZ oscillation line at 8.4 T. Except for this blurred region, full equilibration is consistently observed. To further examine the thermal effects, we measured the resistance of the $P_pP_nP_p$ region at varying temperatures. As temperature increases, thermal noise suppresses the QH effect, leading to a reduction in the clarity of the plateaus. Interestingly, the fluctuations of plateaus decrease with increasing temperature. This behavior is attributed to the disorder effects, which are dominant at low temperature and within Chern insulating regions. Therefore, while disorder-induced fluctuations could be reduced at higher temperatures, the weakening of QH edge channels result in narrower and less distinct plateaus, as shown in **Fig. S11(b) and (d)**. We conclude that the $P_pP_nP_p$ junction exhibits consistent behaviors under varying magnetic fields and temperatures.

The three distinct models are analyzed in Section 3.D using the Landauer-Butikker formalism. The full equilibration case has already been discussed, while the other two cases - selective equilibration and counter propagating edges- are evaluated by introducing additional preserved QH edges and counter propagating channels along the physical edges of the device, as shown in Fig. 4. The number of preserved QH edge channels in the local and global region are denoted as the $N_{local}$ and $N_{global}$, respectively, and the number of counter propagating edges as the N.

For the selective equilibration case, the calculated resistances are:

$$R_{xx} = \frac{h}{e^2}\left[\frac{1}{|v_{global}| - N_{global}} + \frac{1}{|v_{local}| - N_{local}} + \frac{1}{|v_{global}|(|v_{global}| - N_{global})}\right],$$

$$R_{D1} = \frac{h}{e^2}\left[\frac{2}{|v_{global}| - N_{global}} + \frac{1}{|v_{local}| - N_{local}}\right]$$

For the counter-propagating edge case, the corresponding resistances are:

$$R_{xx} = \frac{h}{e^2}\left[\frac{1}{|v_{global}|} + \frac{1}{|v_{local}| + N}\right], R_{D1} = \frac{h}{e^2}\left[\frac{2}{|v_{global}|} + \frac{1}{|v_{local}| + N}\right]$$

In the selective equilibration model, we set $N_{global} = N_{local} = 1$, as shown in Fig. 4(b). As the values of $N_{local}$ and $N_{global}$ increase, both $R_{xx}$ and $R_{D1}$ increase correspondingly, approaching the insulating state. In the counter propagating edge case, the hBN-aligned graphene band structure allows for a large number of counter propagating channels, consistent with the limit $N = \infty$. Based on these calculations, we obtain the modeled $R_{xx}$ and $R_{D1}$ for the three distinct cases, as shown in **Fig. 4(d) and (e)**.

## IV. Helical state in the local area

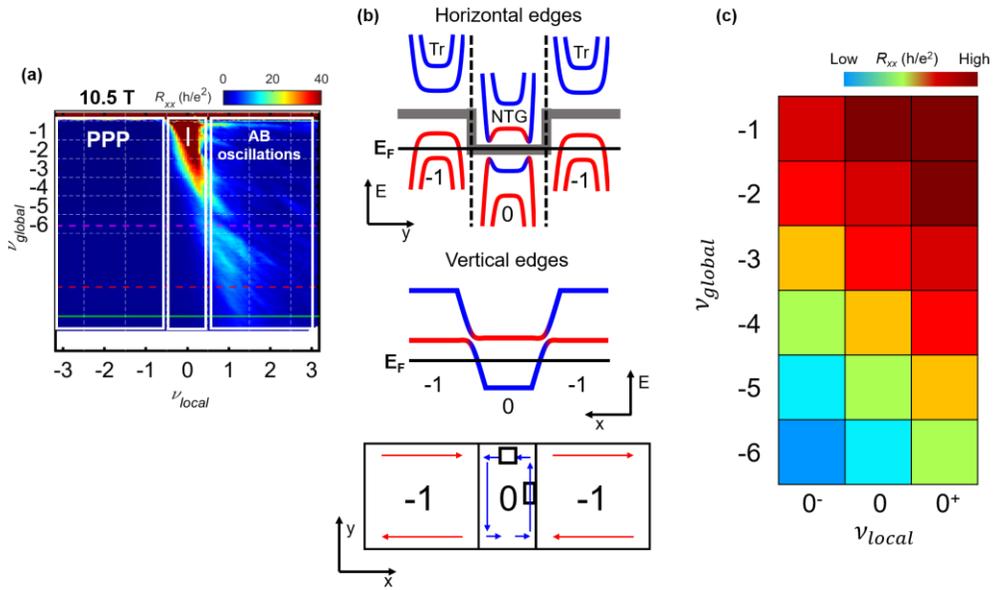

**Fig. S12.** Non-trivial edge effect in $\nu_{local} = 0$ (a) The Rxx graph in 10.5T near the $\nu_{local} = 0$. The insulating region (I) represents the resistance reducing according to global filling factors. (b) The schematic about the formation of non-trivial edges. The partial edge channel exhibits near the corner due to the helical state which has a surface gap. (c) The resistance map of the I region with non-trivial edge channels.

The insulating phase at $\nu_{local} = 0$ (the vertical resistive line) unexpectedly disappears (see also **Fig. S14** at a different magnetic field in more detail). We find that the insulating state at FDP (the vertical insulating area at $\nu_{local} = 0$) in the top-gated area is influenced by the doping level of the global area.

As shown in **Fig. S12(a)**, the highly-resistive region at $\nu_{local} = 0$ deforms to appear as a triangular shape, where the resistance decreases with increasing P-doping. In contrast, in another device measured at 14 T and 40 mK, the triangular-shaped resistance region is absent (see **Fig. S13**; see also **Fig. S14** for the resistance maps in other magnetic fields). We attribute the formation of the triangular-shape resistance feature near the $\nu_{local} = 0$ to the emergence of helical states in the locally gated region [38–40].

The QH edge channels are defined by their spatial positions, consisting of horizontal and vertical edges, as shown in **Fig S12(b)**. Under an optimal helical state, QH edge channels are consistently established, leading to the ET. However, a highly insulating regime emerges at low global doping. To account for this behavior, we propose that the helical state is accompanied by a surface gap [39,40]. Under this condition, the high resistance observed at low global doping and the reduced resistance at high global doping can be attributed to the presence of non-trivial edge states that form near the corners of the locally gated area. These non-trivial edge channels can be interpreted as a surface state emerging at the corner where the band diagram transitions from vertical edges to horizontal edges, as shown in **Fig 12(b)**.

When the Fermi level lies within the surface gap and the global area exhibits low P-doping, non-trivial edge states near the corners of the locally gated region are suppressed, resulting in high resistance, as shown in **Fig. 12(a)**. In contrast, when the global area is tuned toward a higher P-doping, the effect of non-trivial edge states becomes more pronounced, widening the intermediate region between vertical and horizontal edges and reducing the overall resistance. Therefore, the resistance decreases with increasing local and global P-doping, generating the triangular-shaped resistance modulation, as shown in **Fig 12(c)**.

In the case of not having the surface states, such as a conventional chiral state, the device would show high resistance when the Fermi level of the local area reaches the FDP, where a band gap opens due to degeneracy lifting. The vertical FDP region is observed in another device in 14T (**Fig. S13**). However, in this device, the triangular-shaped resistance modulation is absent. We attribute this difference to a screening effect caused by the relatively thin top hBN layer. Previous studies have shown that the helical state in monolayer graphene originated from the sequence of degeneracy lifting between spin and valley [38–40]. When the Coulomb interactions are suppressed by applying an in-plane direction magnetic field or by introducing a screening layer through an insulating layer with thin or high-dielectric constant, the canted antiferromagnetic or ferromagnetic phase can emerge in zLL of graphene, representing helical and helical with surface gap band diagram [39–42].

In our device, we estimate that the thin top hBN layer (11 nm) generates the screening effect in the local area, revealing the helical state. This helical state interacts with the global area, where the chiral state forms due to the thicker bottom hBN (35 nm). In contrast, in another device with a thicker top hBN (15 nm) and bottom $SiO_2$ layers (285nm), the stronger Coulomb interactions inhibits the formation of helical state in both the local and global area, revealing straight-shaped resistance modulation. Therefore, we attribute the observed resistance variations to the interaction between chiral and helical states mediated by non-trivial edge channels emerging in the locally gated area.

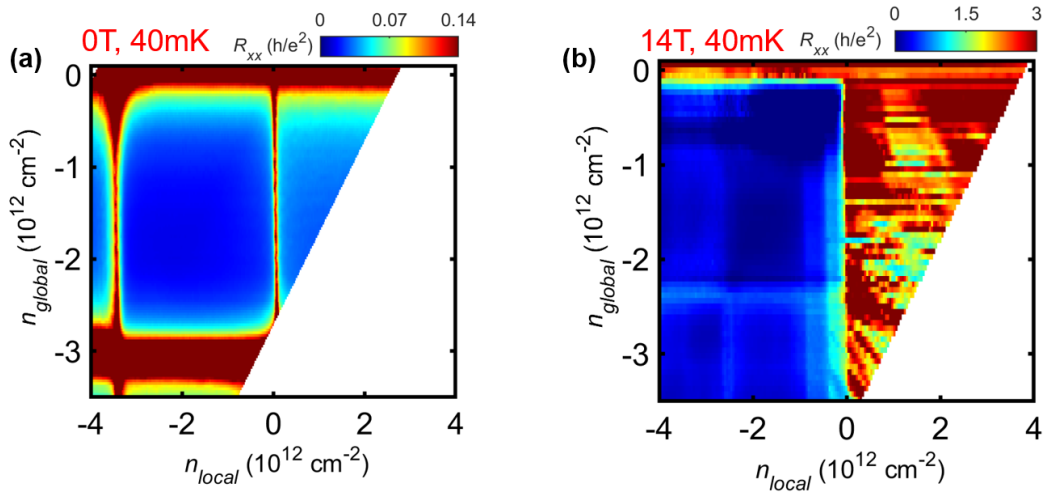

**Fig. S13** Resistance modification upon local and global doping. (a), (b) the resistance variation upon local and global doping in 0 T and 14 T. The FDP line is survived and represents the straight line in the 14 T.

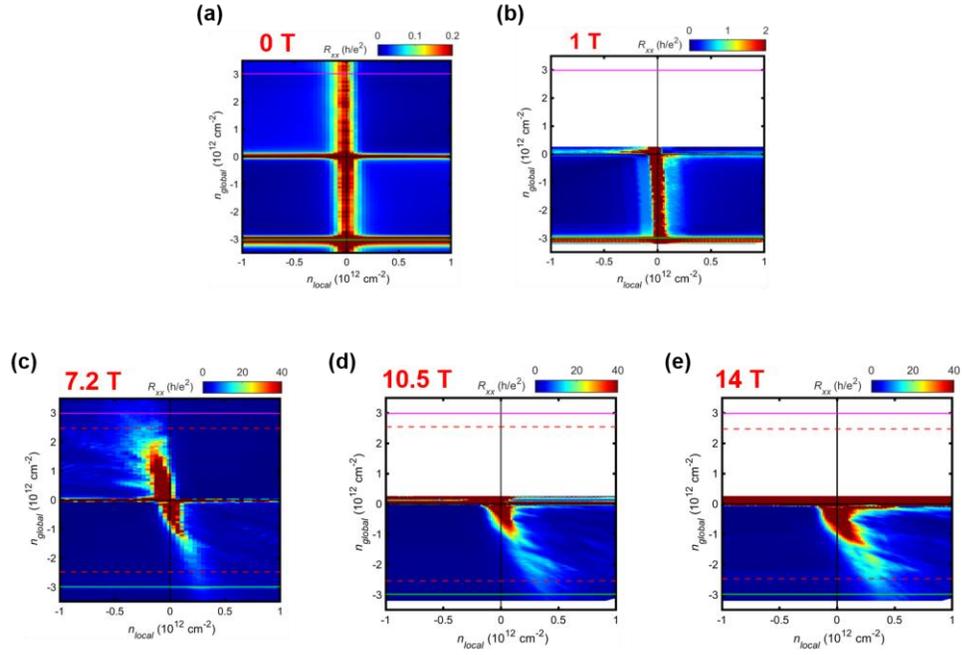

**Fig. S14** Slope of the top-gated area's FDP upon the magnetic fields. (a), (b), (c), (d), (e), and (f) represent the zoom of the top-gated area's FDP. The magnetic fields are 0, 1, 7.2, 10.5, and 14 T.